\documentclass[floatfix,aps,prd,showpacs,amsmath,amssymb,nofootinbib,preprintnumbers]{revtex4}

\begin{document}

\title{Trans-Planckian wimpzillas}

\author{Edward W. Kolb}
\email{Rocky.Kolb@uchicago.edu}
\affiliation{ Department of Astronomy and Astrophysics, Enrico Fermi Institute,
	and Kavli Institute for Cosmological Physics, The
        University of Chicago, Chicago, Illinois, 60637, USA}

\author{A. A. Starobinsky}
\email{alstar@landau.ac.ru}
\affiliation{Landau Institute for Theoretical Physics, Moscow, 119334, Russia}

\author{I. I. Tkachev}
\email{Igor.Tkachev@cern.ch}
\affiliation{Theory Division, CERN, CH-1211, Geneva 23, 
	Switzerland}

\date{\today}

\begin{abstract} 

Two previously proposed conjectures---gravitational trans-Planckian particle 
creation in the expanding universe, and the existence of ultra-heavy stable 
particles with masses up to the Planck scale (wimpzillas)---are combined  in a
proposal for trans-Planckian  particle creation of wimpzillas.  This new
scenario  leads to a huge enhancement in their production compared to
mechanisms put forward earlier. As a result, it requires the trans-Planckian
particle creation  parameter to be rather small to avoid overproduction of such
particles, much less than that is required for observable effects in the 
primordial perturbation spectrum. This ensures also that wimpzillas are  mainly
created at the end of primordial inflation. Conditions under which 
trans-Planckian wimpzillas can constitute the present dark matter are 
determined.

\end{abstract}


\maketitle 

\section{Introduction}

Creation of pairs of all types of particle and antiparticles by a strong
external gravitational field is a direct analogue of electron--positron 
creation in a strong electromagnetic field--an unambiguous 
prediction of quantum electrodynamics. The former effect also has a solid 
field-theoretical basis. Its relevance in cosmology was recognized  
by Schr\"{o}dinger as early as 1939 \cite{alarming}. In his paper, 
``The proper vibrations of the expanding universe,'' Schr\"{o}dinger 
discussed what he referred to as the  ``alarming phenomenon'' of particle 
creation in an expanding universe. The use of the word `alarming' suggests 
that emergence of particles from the quantum  vacuum simply due to the 
expansion of the universe seemed to Schr\"{o}dinger at the time to
signify some internal inconsistency of quantum field theory. However,
such concerns completely disappeared after the construction of a
rigorous theory for gravitational creation of particles and the
energy-momentum tensor of quantum fields in cosmological backgrounds,
beginning with the pioneering papers in Refs.\ \cite{ChT68} (the de Sitter
background), \cite{P68,GM70} (a Friedmann-Robertson-Walker (FRW) 
background), \cite{ZS72} (an anisotropic homogeneous background and the
energy-momentum tensor), \cite{G75} (gravitons in a FRW background),
and others. For many years after that, this theory was considered as
a purely theoretical exercise, far removed from any practical applications.
Now, observations verifying the direct consequences of this effect have 
become one of the main topics in  experimental and theoretical
cosmology, since gravitational particle  creation in the expanding
universe during a primordial inflationary (de Sitter) stage  serves as the
physical mechanism for the generation of the observed inhomogeneities in the
matter density and the cosmic microwave background (CMB) radiation temperature
\cite{sc} (along with a predicted, but yet unobserved, primordial gravitational
wave background \cite{gw}). 

The usual calculations of density perturbations assuming the minimal possible
choice for the initial quantum state for perturbations (namely, adiabatic
vacuum initial conditions for each Fourier mode), have led to predictions that
have been confirmed by observations.  This is considered as one of the most
remarkable successes of the whole inflationary scenario. Moreover, it should be
noted that even alternatives like the Pre-Big-Bang \cite{pre-big-bang}
or ekpyrotic
\cite{ekpyrotic} scenarios, in which there is no inflationary stage, still use
the phenomenon of particle creation in the expanding universe to produce the
observed perturbations. So the relevance of particle creation is not tied to a
specific cosmological scenario.

However, the phenomenon may have even more richness:     
Consider the hypothesis of the existence of supermassive particles with a
rest mass $m > 10^{10}$ GeV and a lifetime exceeding the age of the universe.
These hypothetical particles were dubbed wimpzillas in Ref.\ \cite{ckr}. Here
the question of how to produce them in the early universe arises once more. One
possibility is the standard gravitational particle creation scenario first
studied in Refs.\ \cite{ckr,kt}. Then, production at the end of inflation
appears to be the most efficient for this aim. In such a scenario,
production of 
particles of mass in excess of $H_I$ is strongly suppressed. Since we know that
$H_I \lesssim 10^{14}$ GeV from the upper limit on the contribution of
primordial gravitational waves to the measured CMB  fluctuations, wimpzillas
with masses exceeding $10^{14}$ GeV cannot be produced by this mechanism. An
additional possibility for production of higher mass  particles  is provided by
the preheating process \cite{preh},  {\it i.e.,} the rapid creation  of massive
particles by inflaton oscillations after the end of inflation in the regime of
a broad parametric resonance. Here, one may expect creation of particles with
masses up to $10^{16}$ GeV \cite{ckr}, but not more.\footnote{Here we speak
about boson production. Mechanisms of fermion  production during preheating
\cite{gprt}, or during inflation through direct coupling of a wimpzilla to an
inflaton \cite{ckrt}, are capable of producing fermions with a rest mass as
large as the Planck mass.  The vast difference between fermion and boson cases
reflects the non-gravitational origin of these kind of mechanisms.}

Now the question arises if it may be possible to create even more wimpzillas 
than the known mechanisms listed above permit. We put forward a new idea to 
use a (hypothetical, of course) mechanism of {\it trans-Planckian particle 
creation} (TPPC) for this aim. The TPPC effect during inflation was first 
proposed in Ref.\ \cite{BM01}. Though the initial emphasis in TPPC
investigations 
concentrated on the uniqueness of the inflationary predictions for
perturbations in standard quantum field theory (QFT), it  soon
became clear that TPPC is a new physical effect that exists not only during
inflation, but for all of cosmic
history (if it exists at all), and it requires breaking some postulates 
of  standard QFT for its existence (in particular, local Lorentz 
invariance). Ref.\ \cite{MR06} contains an extensive 
list of further references on this topic, including various microscopic 
mechanisms that might lead to such an effect. We will not discuss 
possible mechanisms of the TPPC effect in this paper. Rather, we will 
restrict ourselves to a purely phenomenological description of its
outcome in terms of `out' parameters defined after the TPPC effect 
ceases,
and we will use its observational consequences to place new and 
severe restrictions on the possible strength of the TPPC effect.

In Sec.\ II the TPPC description is introduced and several subtle points about
the TPPC effect are discussed, in particular, why the TPPC effect  generically
{\em may not} be reduced to standard particle creation  from a non-vacuum
initial quantum state, and what restricts the possible amount of TPPC. These
points are necessary both for the strict definition of what we mean by TPPC,
and since some confusion still seems to exist in the literature regarding this
topic. A reader interested in the direct application to observational effects
may skip this section in the first reading. In Sec.\ III, TPPC creation of
wimpzillas is calculated and new restrictions of the TPPC effect are obtained
from it. Final conclusions are presented in Sec.\ IV.

\section{TPPC versus standard particle creation from a non-vacuum initial
state}

As has been pointed out above, standard QFT predictions for particle creation
in curved space-time, including the inflationary predictions for
perturbations, are based on an adiabatic in-vacuum for {\em all} modes with a 
sufficiently large momentum. Of course, this is crucial for the uniqueness
of the final result. Let us emphasize that the choice of an initial
quantum  state is a physical, not a technical, problem. The whole discussion 
of  TPPC historically began from the reconsideration of the well known
fact that for any Fourier mode  ${\bf k}~(k=|{\bf k}|)$ of small 
inhomogeneous perturbations of a FRW background with a scale factor $a(t)$, 
the physical momentum $p=k/a(t)$ was once very large at sufficiently early 
times during the 
expansion of the universe and may have greatly exceeded the Planck mass 
$M_{Pl}=1/\sqrt G$ ($\hbar = c = 1$ is assumed throughout the paper). Thus, 
observable modes of density and temperature fluctuations emerged from a
`trans-Planckian' region. 

However, this does not result in anything dangerous
for a Lorentz-invariant theory and does not preclude the unambiguous
determination of the vacuum state so long as $\omega^2(p) - p^2$ remains  much
less than $M_{Pl}^2$, where $\omega$ is the physical frequency. Let us
recall that exactly the same trans-Planckian problem arises formally
in the calculation of
electron-positron pair creation in a constant electric field ${\bf E}_0$ if 
the gauge ${\bf A} = -{\bf E}_0t$ is used.  However, it can be easily verified
(see, {\em e.g.,} Ref.\ \cite{N70}) that by using the adiabatic  vacuum as an
initial condition for all Fourier modes at $t\to - \infty$ and integrating
over all momenta $p$, in spite of the unlimited growth  of particle energy in
this limit, one correctly reproduces the Heisenberg--Euler action (more
exactly, its analytical continuation to the electric field case) obtained by
fully covariant methods. Thus, using the canonical approach to QFT we are even 
{\em obliged} to consider field modes with an arbitrarily high momentum $p$,
but in the adiabatic vacuum state, in order to obtain correct results: 
no momentum cutoff is permitted.     

Still, in the cosmological case, as well in the toy model with an electric 
field, the initial state may well be a non-vacuum state and contain some 
particles (well defined up to exponentially small terms so long as
their momenta or rest mass exceed the
Hubble parameter $H\equiv \dot a/a$). So, the physical question concerns 
their origin. One possibility, which does not require a change of basic 
physical laws, is that these particles were produced either before the 
beginning of inflation or during some previous stage(s) of inflation with a 
higher curvature. Indeed, {\em e.g.,} in a model with two stages of inflation 
divided by a matter-dominated or radiation-dominated period, some modes 
enter the second inflationary stage in a non-vacuum state  \cite{PS92}. 
However, the average number of particles and their energy density in the
initial state taken at some moment $t=t_0$ should be finite and
not too large to avoid an excessive back-reaction incompatible with the 
assumed behavior of a FRW background:
\begin{equation}
\langle\rho_{part}\rangle= \frac{g_s}{(2\pi)^3a^4}\int_{k=aH}^{\infty} d^3k~
\omega(k)\langle n({\bf k})\rangle ~\lesssim H^2M_{Pl}^2
\label{initial}
\end{equation} 
at $t=t_0$ ($g_s$ is the statistical weight).\footnote{In generalizations of
the general theory of relativity, such as scalar-tensor gravity, $F(R)$ theory
where $R$ is the Ricci scalar, brane gravity, {\it etc.,} $G_{eff} ^{-1}$
(depending on $H$, $\rho$, and other quantities) replaces $M_{Pl}^2$ on the
right-hand side of Eq.\ (\ref{initial}).} This requires that $k^3\langle n({\bf
k})\rangle \to 0$ as $k\to \infty$ for any kind of quantum state. As a result,
the effect of a physically admissible non-vacuum initial state on the
perturbation power spectrum is very transient in $k$-space. Deviations of
spectra from the vacuum result, proportional to $\sqrt{\langle n({\bf
k})\rangle}$, must become negligible for large $k$ (late Hubble radius crossing
times  $t-t_0\gg H^{-1}$ during inflation). For this conclusion to be valid, it
is not necessary that the initial state of  each mode decay to vacuum (due to,
{\it e.g.,} some particle interactions) and it may even remain as it
is. This shows, in particular, that the slowness of relaxation pointed
out recently in Ref.\ \cite{AP07} does not present any obstacle for the 
unambiguity of predictions of the inflationary scenario. The
self-consistency condition (\ref{initial}) is crucial for all discussions 
of which initial states are typical. It plays the same role as the fixing 
of the energy for a microcanonical ensemble or temperature for a canonical one 
in statistical physics. Without it, of course, particles of arbitrary
high energy would be typical at all times including at present. 

In addition, creation from a non-vacuum initial state results
in a drastic deviation of the initial power spectrum of density perturbations
from the approximately flat (Harrison-Zel'dovich) spectrum that is easily
distinguishable. Within present observational bounds, such behavior is
possible only for scales close to the present Hubble radius; see Refs.\
\cite{MR06,feat} for the recent comparison of different local features in the
power spectrum with the 3-year WMAP data.

Thus, significant (over a wide interval of scales) corrections to the standard
inflationary predictions are possible only with a hypothesis of the existence
of a new effect---{\em trans-Planckian particle creation}. 
Observational signatures of this effect give us tools to study and constrain
physics in the trans-Planckian regime, which explains the recent attention
devoted to this subject. The main feature distinguishing TPPC from standard
gravitational particle creation is that the characteristic energy of created
particles and antiparticles (the symmetry between matter and antimatter is
still respected) at the moment of their creation is not $E\sim H$ (for a
rest mass $m\ll H$) but $E\sim \Lambda$, where $\Lambda$ is some new scale
that should necessarily be connected with some kind of (possibly soft) Lorentz
invariance breaking, otherwise such a process is impossible.\footnote{Unless
explicitly stated, we will assume $H\ll\Lambda$.} The natural candidate for
$\Lambda$ is the Planck mass; however, there are other possible
scales: the superstring scale, the scale associated with duality (a minimum
length scale), scales associated with extra dimensions, and so on. In these
cases, one expects that $10^{-3}M_{Pl}\lesssim\Lambda \lesssim M_{Pl}$.

Soon after the initial proposal, it was emphasized in Ref.\ \cite{S01} that if
the TPPC effect exists at all, it should be ``everlasting:'' {\em i.e.,} it
should not be restricted to only the de Sitter background, but it should occur
during all periods of the universe's expansion up to the present time. Indeed,
the main reason for the possible existence of the TPPC effect---continuous
drift of all Fourier modes  from the trans-Planckian region of momenta to the
sub-Planckian one---remains the same for any kind of expansion. Of course, the
TPPC effect  should become  weaker for a smaller curvature (a
curvature-independent effect is immediately excluded by its absence at present
\cite{S01}). 

Since the TPPC phenomenon produces particles of energy $\Lambda$, ultra-high
energy particles should be created long after primordial inflation (and even
now!). The dependence of TPPC on $H$ is crucial for determining the moment
of time when created particles have the most important effect on any
observable. Furthermore, one could hardly expect that it is possible to break
Lorentz invariance while keeping conformal invariance intact.  Therefore, TPPC
should be ``democratic,'' and the creation of photons, neutrinos,  electrons,
{\em etc.}, should be possible, in addition to creation of scalar (density)
perturbations and gravitons (tensor perturbations). This was used by two of us
in Ref.\ \cite{ST02} to obtain strong restrictions on the TPPC effect using
observed limits on the flux of the diffuse X-ray background. 

Note that the TPPC effect, which corresponds to a non-standard contribution to 
the imaginary part of the (off-shell) graviton propagator, should be 
distinguished from vacuum polarization effects (corrections to the real part 
of the propagator). The latter ones certainly exist and lead, in particular, 
to $H_I^2/M_{Pl}^2$  corrections to the standard inflationary predictions 
(here $H_I$ is the value  of $H$ during inflation--more exactly, during the
last  60 $e$-folds of it).\footnote{Effects of the same order were considered
in Refs.\ \cite{pol} and other papers.} However, this only results in an 
effective renormalization of $H_I$ and other parameters defining the primordial
spectra. On the other hand, the TPPC effect is a much bolder hypothesis, and it
may well not exist at all. But if it exists, it leads to unique consequences
such  as the generation of super-high-energy cosmic rays at the present time
\cite{ST02}  and oscillations in the primordial power spectra of scalar and
tensor  perturbations with amplitudes not decreasing with the growth of $k$
\cite{D02}.

Following Refs.\ \cite{S01,ST02,D02,BGV03}, for a quantum field $\phi$, the
TPPC effect can be completely, though only phenomenologically, described in
terms of the Bogoliubov coefficients $\alpha_k$ and $\beta_k$ (the Bogoliubov
coefficients satisfy $|\alpha_k|^2 - |\beta_k|^2 = 1$) in the expression for
the time-dependent part $\phi_k(t)$  of the mode wave functions multiplying the
Fock annihilation operator $\widehat{a}_k$ in the Heisenberg representation 
valid during the WKB regime $p=k/a(t)\gg H$, but after the mode has reached the
sub-Planckian region $p\ll \Lambda$:
\begin{equation}
\phi_k=\frac{1}{\sqrt{2\omega a^3}}\left(\alpha_k e^{-i\int \omega dt} +
\beta_k e^{i\int \omega dt}\right) ~,
\label{alphabeta}
\end{equation}
where $\omega=\sqrt{p^2+m^2}$ and the rest mass $m$ are assumed to be much less
than $\Lambda$. Though only the leading terms (the zeroth- and first-order
ones) of the WKB solutions are explicitly written in this formula, the
full WKB series is  there.  

The `final' condition (\ref{alphabeta}) (final in the sense that TPPC  ceases)
corresponds to the mode ${\bf k}$ being in a pure  squeezed quantum state. This
assumption is taken by analogy with the usual  particle creation from vacuum
where the final state is a squeezed one, too.  However, since we need the
average particle number $\langle n({\bf k})\rangle \equiv |\beta_k|^2$ only for
the calculation of TPPC of wimpzillas (see Sec.\ III below), it is
straightforward to generalize the condition (\ref{alphabeta}) to an arbitrary
pure quantum state having the same $\langle n({\bf k})\rangle$. In principle,
further generalization to the density matrix description is also possible and
does not present any problem. However, since usual particle creation both in
cosmology and black hole physics does not result in the local loss of
coherence, {\em i.e.,} in the transformation of a pure state into a mixed one,
it is natural to suppose (at least, at the first step) that the TPPC effect
does not lead to it either.   

The surface $p^2=\Lambda^2$ was dubbed the `new-physics hypersurface' in Ref.\
\cite{BGV03} (see Ref.\ \cite{EKP05} for a comparison of this approach to other
ones). Corrections to the perturbation spectra produced during inflation are 
proportional to $\beta_k$, while the number of created particles is 
$n_k=|\beta_k|^2$.  As was pointed out above, there is no TPPC effect in flat
space-time, so as expected, $\beta_k=0$ for $H=0$. The main hypothesis assumed
in all studies of  TPPC is that $\beta_k$ is not exponentially suppressed, but
only power-law suppressed for $H \ll \Lambda$, where $H$ is taken at the moment
$k=a\Lambda$ when the  given mode crosses the new physics hypersurface
(otherwise, the effect would not be of practical interest).  Thus,
$\left|\beta_k\right|$ may be modeled as
\begin{equation}
\left|\beta_k\right| = b \left( \frac{H_k}{\Lambda} 
\right)^{\gamma}\ll 1,~~~~\textrm{with } \gamma > 0~,
\label{betah}
\end{equation}
where $H_k$ is the value of the expansion rate when a mode with comoving
momentum $k$ crossed the trans-Planckian region: $H_k \equiv H(k=a\Lambda)$.

Of course, the use of only one time-dependent parameter $H_k$ is an
over-simplification. Really, from general covariance one should expect
that for a FRW background $\beta_k$ depends not only on $H$ but also
on $\dot H/H^2$ in such combinations that the actual dependence is on 
the generally covariant quantities $R\equiv R^{\mu}_{~\mu}$ and 
$R_{\mu\nu}R^{\mu\nu}$. Moreover, this dependence is most probably 
non-local since even the usual particle creation effect is always 
non-local. Also, the answer should contain the step function 
$\theta(H)$ since the very TPPC effect may come into existence only due to
the expansion of the universe. Thus, the formula (\ref{betah}) is a crude
approximation which, however, becomes a rather good one during 
slow-roll inflation since $R_{\mu\nu}$ is approximately constant and 
uniquely related to $R$ there.

In principle, the parameter $\gamma$ need not even be an integer. In
particular, just such  a case is realized in the concrete model proposed in
Ref.\ \cite{LR05} where four-dimensional Lorentz invariance is softly violated
due to brane-world effects (see Ref.\ \cite{CEG01} for earlier works in a
similar direction). However, mostly integer values of $\gamma$ were considered
previously. The condition of the absence of an excessive back-reaction for
TPPC requires
\begin{equation}
\left|\beta_k\right| \lesssim H_k M_{Pl}\Lambda^{-2} ~ ,
\label{back}
\end{equation}
which is valid both during inflation \cite{T00} and after it \cite{S01} (see
also \cite{AL03,AMM05}). It is obtained under the assumption that
created particles
appear just at the  new-physics hypersurface $p=\Lambda$, and do not exist
before that. The  comparison of Eq.\ (\ref{back}) with the corresponding
condition for the standard quantum field theory of Eq.\ (\ref{initial}) (with
$\langle n({\bf k})\rangle=|\beta_k|^2$) clearly  shows the profound difference
between the two cases: First, integration over $k$ from the Lorentz
non-invariant scale $k=a\Lambda$ up to infinity is completely omitted and
modeled as a boundary term at $k=a\Lambda$; second, this boundary term is
estimated at different moments of time for different $k$, in contrast to Eq.\
(\ref{initial}) imposed at the same $t=t_0$ for all modes. 

One may ask how the mode wave function (\ref{alphabeta}) would look at 
earlier times (in particular, if one wishes to define initial conditions 
for all modes ${\bf k}$ at the same moment of time). The answer is that
in standard QFT where TPPC is absent, it has essentially
the same form as (2) with constant $\alpha_k$ and $\beta_k$ since it
remains in the WKB regime.  For a non-standard QFT admitting TPPC, the
answer is, of course, strongly model dependent and generically not known at
all. Then, however, condition (\ref{initial}) ceases
to be valid soon in the past due to the decrease of $a(t)$ (especially
rapid during inflation). In particular, even the partial contribution
to the integral over $d^3k$ in (\ref{initial}) from modes having 
$k=a\Lambda$ at the end of inflaton 
increases by more than $e^{240}\approx 10^{104}$ times when
shifted  to an initial moment (presumably the same for all modes)
taken slightly earlier than the moment when a comoving scale equal to the 
present Hubble radius first crossed the Hubble radius during inflation. So, 
in order not
to violate (\ref{initial}), either $\gamma$ should be very large or
$b$ should be exponentially small if we wish to obtain the result 
(\ref{betah}) using standard particle creation from a non-vacuum 
initial state. Moreover, not only the power-law behaviour (\ref{betah}) 
with an arbitary $\gamma$ but any phenomenological outcome of TCCP 
not being {\em exactly} zero for a sufficiently small $H$ {\em may 
not} result from any non-vacuum initial state in standard QFT if
the de Sitter stage beginning in our Universe now is stable since 
$|\beta_k|\to const$ at $k\to \infty$ for that state.  
   
This shows that the condition (\ref{initial}) yields a quantitative 
criterion to distinguish genuine TPPC from that which may be explained
without it. Note that this conclusion is opposite to the main
statement of Ref.\ \cite{SP05} about the impossibility to discriminate 
between the usual and trans-Planckian particle creation using observational 
data. Though for a test QFT in an external gravitational field, any 
final outcome of TPPC like (\ref{alphabeta}) may be formally expressed 
as the usual creation from a non-vacuum initial state taken at some earlier
time, this construction becomes inconsistent when the 
energy-momentum tensor of particles is taken into account. In other
words, we {\em may not} simply employ standard QFT and think that 
particles  really exist if  condition (\ref{initial}) is
violated. Thus, particles created by TPPC, with their energy and momentum 
in the usual sense, should come to existence at some later time only,
not at the initial hypersurface with equal time for all modes ${\bf k}$.  
  
Inserting Eq.\ (\ref{betah}) into Eq.\ (\ref{back}) we see that for $\gamma \ge
1$, Eq.\ (\ref{back}) is satisfied for all $H\le \Lambda$ if $b\le
M_{Pl}/\Lambda$.  Thus, in the $\gamma\ge 1$ case, Eq.\ (\ref{back}) is not very
restrictive.   However, if $\gamma < 1$, then the inequality in Eq.\
(\ref{back}) is  violated for sufficiently small $H$ regardless of $b$.

The case $\gamma=1$ is required to obtain noticeable corrections to the 
primordial spectra generated during inflation. The above mentioned results of 
Ref.\ \cite{ST02} show that the $\gamma=1$ TPPC effect must be strongly
suppressed for usual elementary particles: $b<10^{-6}M_{Pl}/\Lambda$.  This
result leads to the absence of any noticeable features in the CMB temperature
anisotropy. On the other hand, ultra-high energy particles created in this way
at the present time still can be seen through an excess of cosmic rays at
energies above the Greizen-Zatsepin-Kuzmin limit (if such an excess will prove
to exist). 

In Eq.\ (\ref{betah}), it is assumed that $\Lambda$ is the {\em minimal}
energy  scale connected to `new physics' and that this expression is valid for
all  energy scales $\omega \ll \Lambda$. Inflation is supposed to occur below
this scale also: $H_I\ll \Lambda$. To avoid the upper limit of Ref.\
\cite{ST02} keeping the possibility to detect some TPPC features in CMB
fluctuations, a more complicated model with {\em two} scales $\Lambda$ and
$\Lambda_1 \ll \Lambda$ was proposed in Ref.\  \cite{BM05}, where Eq.\
(\ref{betah}) with $\gamma=1$ is valid for $\Lambda_1\ll H \ll\Lambda$ and
inflation occurs also in this energy interval. On the other hand, $|\beta_k|$
is strongly suppressed for $k\ll \Lambda_1$. Of course, assuming inflation to
occur above any new-physics scale, especially that related to some kind of the
Lorentz invariance breaking, strongly undermines the power and beauty of the
standard  inflationary calculations based just on the assumption of the {\em
absence} of any radically new physics up to the scale of inflation. In
particular, if  Lorentz invariance is not valid during inflation, then even the
necessity to invoke inflation to explain causal connections in the observed
part of the universe may well be called into question. Fortunately, present
observations do not require us to go so far: the direct search of superimposed
oscillations of potentially trans-Planckian origin in the power spectrum of 
density perturbations using  the 3-year WMAP CMB data \cite{MR06} (see also 
Ref.\ \cite{RZM06}) does not give any  statistically significant evidence for 
them. All this puts the existence of  the TPPC effect with $\gamma \le 1$ under
serious question. However, larger  values of $\gamma$ are not excluded. In the
next Section, we show that TPPC of wimpzillas presents a new possibility 
to probe that range of $\gamma$.

\section{TPPC of wimpzillas}

Let us combine the TPPC and wimpzilla ideas and study the cosmological 
implications of the creation of wimpzillas by the expansion of the universe due
to trans-Planckian effects. Due to the universality of gravitational
interactions, TPPC, if exists at all, should occur for all types of particles
and for all times. Moreover, trans-Planckian production of wimpzillas is not
suppressed as long as their rest mass satisfies $m_X < \Lambda$. Thus, the TPPC
effect opens a possibility for a new and more effective way to produce
super-heavy particles with masses up to $M_{Pl}$ (if $\Lambda \sim M_{Pl}$). In
turn, supermassive particles present a new possibility for TPPC to reveal its
properties in a different regime: since wimpzillas are not thermalized, the
present abundance of particles of mass in excess of $H_I$ should reflect the
TPPC properties, in particular, the value of $\gamma$.

The number density of created $X$ (wimpzilla) particles
is given by the expression \begin{equation} n_X = \frac{1}{2\pi^2 a^3}
\int_{0}^{a\Lambda} dk \, k^2 \left|\beta_k\right|^2 ~, \label{nxc}
\end{equation} where the upper limit just reflects the assumption that
particles are created with the physical momentum $\Lambda$ at the moment of
time  when $k/a(t)=\Lambda$, {\em i.e.,} at different moments of time for
different modes. After that, their momentum simply is redshifted as the scale
factor $a$ increases.

The next step is to determine the moment when TPPC produces the main effect. 
Let us insert here $|\beta_k|$ from Eq.\ (\ref{betah}). The dependence 
of the expansion rate $H$ on $a$ depends on the cosmic epoch as
\begin{equation}
H \propto \left\{ \begin{array}{ll}
a^0		& \textrm{inflation} 	\\
a^{-3/2} 	& \textrm{matter dominated} 	\\
a^{-2} 		& \textrm{radiation dominated}~ .\end{array} \right. 
\end{equation} 
We see that for any positive  value of $\gamma$, the contribution to the total
number density from particles that  emerged from the trans-Planckian region
during inflation is dominated by those created toward the end of this epoch 
(those created earlier are redhifted away). During the matter-dominated era, 
late-time TPPC dominates if  $\gamma\leq 1$, while if $\gamma>1$, then the 
largest contribution comes from the particles created at the beginning of the 
matter-dominated phase.  During the radiation-dominated era, late-time TPPC
dominates for $\gamma\leq 3/4$, but for $\gamma > 3/4$ early-time created 
particles make the main contribution to $n_X$. 

In view of the discussion above, let us turn to the case $\gamma > 1$ from here
on. Then the integral in Eq.\ (\ref{nxc}) is dominated by the contribution
from the end of inflation. For the accuracy needed, it is sufficient to put
$H=H_I=const.$ in this region. Then the $X$-number density after inflation
would be
\begin{equation}
n_X =\frac{b^2}{6\pi^2}\left(\frac{H_I}{\Lambda} \right)^{2\gamma} \Lambda^3
\left(\frac{a_{EI}}{a}\right)^3 ~,
\end{equation}
where the $EI$ subscript denotes the end of inflation.

Now the development depends a bit on the evolution of the universe between the
end of inflation and ``reheating,'' the beginning of the radiation-dominated
era.  We will assume the maximum temperature of the universe in the
radiation-dominated era was $T_{RH}$. If entropy was conserved between
reheating and today, the ratio of the $X$-number density to the entropy density
$s$ is
\begin{equation}
\frac{n_X}{s} = \frac{b^2}{3\pi}\frac{H_I^{2\gamma-2}}{\Lambda^{2\gamma-3}}
\frac{T_{RH}}{M_{Pl}^2}~ .
\end{equation}
Since it is not very useful to have the result depend on the unknown parameter
$T_{RH}$, we define a reheating efficiency factor $r$ as
\begin{equation}
r = \frac{T_{RH}}{M_{Pl}^{1/2}H_I^{1/2}}~.
\end{equation}
If reheating is very efficient, then $r\sim 1$.  If the extraction of the
inflaton energy density is an inefficient, prolonged affair, then it is
possible that $r\ll 1$. It is also convenient to define a dimensionless 
parameter $\lambda$ relating the trans-Planckian scale $\Lambda$ to the 
Planck scale:
\begin{equation}
\Lambda = \lambda M_{Pl} ~.
\end{equation}
One expects $10^{-3}\lesssim\lambda\lesssim 1$.

In terms of these dimensionless parameters, the ratio of the number density to
the entropy density becomes
\begin{equation}
\frac{n_X}{s} = \frac{1}{3\pi}\frac{rb^2}{\lambda^{2\gamma-3}}
\left(\frac{H_I}{M_{Pl}}\right)^{2\gamma-3/2} ~.
\end{equation}
It is straightforward to convert $n_X/s$ into an expression for $\Omega_Xh^2$,
where as usual $\Omega_X$ is the present ratio of the $X$ energy density to the
critical density and $h$ is Hubble's constant in units of $100 \textrm{ km
s}^{-1}\textrm{ Mpc}^{-1}$:
\begin{equation}
\Omega_Xh^2 =  3\times10^{26} \frac{rb^2}{\lambda^{2\gamma-3}}
\left(\frac{H_I}{M_{Pl}}\right)^{2\gamma-3/2} \frac{m_X}{M_{Pl}} ~.
\end{equation}

In the case $\gamma=2$, to avoid overproduction of wimpzillas, we need:
\begin{equation}
b^2 \times  \frac{m_X}{M_{Pl}} \lesssim
5\times10^{-28}\; \frac{\lambda}{r}\left(\frac{H_I}{M_{Pl}}\right)^{-5/2}  ~.
\label{present}
\end{equation}
Thus, if wimpzillas exist, the strength of the TPPC effect (the coefficient
$b$) should be small even for $\gamma=2$.  Moreover, we see that 
trans-Planckian
effects can produce supermassive dark matter (wimpzillas) as massive as the 
Planck mass in an abundance to result in $\Omega_Xh^2\sim 0.15$, even for 
$H_I$ as low as $10^8$ GeV, corresponding to an energy density during inflation
of about $(10^{13}\,\textrm{GeV})^4$.

On the other hand, in the case of the two-scale trans-Planckian model of 
Ref.\ \cite{BM05} discussed above, we can put $\gamma=1$ in
Eq.\ (\ref{present}) if $H_I\simeq \Lambda_1$, so that there is no TPPC 
creation after inflation. Then 
\begin{equation}
\Omega_Xh^2 = 3\times10^{26}\, rb^2\lambda^{3/2}\left(\frac{\Lambda_1}
{\Lambda}\right)^{1/2}\frac{m_X}{M_{Pl}}~.
\end{equation}
So, even for this model, it is impossible to have noticeable trans-Planckian
effects in CMB fluctuations (which requires $b = {\cal O}(1)$) and the 
existence of wimpzillas at the same time---trans-Planckian creation of the 
latter ones appears to be too strong.

\section{Conclusions}

We have proposed to combine the hypothesizes of the wimpzillas and TPPC
existence and calculate the amount of these superheavy particles
created due to the expansion of the universe. It has been shown that a rather 
weak TPPC effect with $\gamma\simeq 2$ can produce the necessary amount of such 
particles to comprise the present dark matter in the
universe, even if their mass is comparable to the Planck mass. These 
particles are mainly created at the end of inflation. Thus, a new
possibility for the origin of dark matter opens. 

On the other hand, existence of wimpzillas places new upper bounds on
any more fundamental theory predicting TPPC, which are significantly
stronger than those known previously. In particular, it is not possible
to obtain both super-heavy dark matter particles and observable 
corrections to the primordial perturbation spectrum -- TPPC 
cannot bear two fruits at once.

\acknowledgments
 
E.W.K.\ was supported (in part) by NASA grant NAG5-10842. E.W.K. and A.A.S.
would  like to thank the CERN Theory group for hospitality during the beginning
of  this project. A.A.S.\ was also partially supported by the Russian
Foundation  for Basic Research, grant 05-02-17450, and by the Research Program 
``Elementary particles" of the Russian Academy of Sciences.



\end{document}